\documentclass[aps,prl,twocolumn,showpacs]{revtex4}

\usepackage{graphicx}
\graphicspath{{pict/}{./}}

\usepackage{bm}%

\newcommand\pictc[5]{\begin{figure}
                       \centerline{\vspace{-1mm}
 \includegraphics[width=#1\columnwidth,height=0.7\textheight,keepaspectratio]{#3}}
                       \protect\caption{\protect\label{#4} #5}\vspace{-3mm}
                    \end{figure}            }

\newcommand\pict[4][1]{\pictc{#1}{!tb}{#2}{#3}{#4}}
\newcommand\rpict[1]{\ref{#1}}

\newcounter{Fig}

\newcommand{\be}{\begin{equation}}
\newcommand{\ee}{\end{equation}}

\begin{document}
\title{Polychromatic nanofocusing of surface plasmon polaritons}
\author{Wei Liu}
\author{Dragomir N. Neshev}
\author{Andrey E. Miroshnichenko}
\author{Ilya V. Shadrivov}
\author{Yuri S. Kivshar}
\affiliation{Nonlinear Physics Centre, Centre for Ultrahigh-bandwidth Devices for Optical Systems (CUDOS), Research
School of Physics and Engineering, Australian National University,
Canberra, ACT 0200, Australia}

\pacs{
        78.67.-n,   
        78.67.Pt,   
        42.25.Bs    
}

\begin{abstract}
We introduce the concept of polychromatic plasmonics and suggest an broadband plasmonic lens for nanofocusing of surface plasmon polaritons. The lens employs a parabolically modulated metal-dielectric-metal structure. This plasmonic lens has a bandwidth of more than an optical octave thus opening new opportunities for broadband plasmonic applications.
\end{abstract}
\maketitle

The field of plasmonics experiences an explosive growth recently due to the ability of plasmonic components to confine light
down to the nanoscale. Various plasmonic structures with
miniaturization scales comparable to those of modern semiconductor
electronics have been designed to realize different types of light
waveguiding and control~\cite{Cao2009, Zayats2005, Gramotnev2010,
Lee2010, Liu2010, Huidobro2010, Verslegers2009}. Despite the boom
in research one can identify two major challenges for further applications of plasmonic devices: high propagation loss and strong wavelength dispersion. Different approaches to combat
optical losses in metals have been explored, including incorporation
of gain~\cite{Noginov2008} or nanofocusing~\cite{Stockman2004,Davoyan2010}. However, the concept
of broadband control and nanofocusing of {\em polychromatic surface plasmon polaritons} (SPPs) remains practically unattainable.

In this Letter, we introduce the concept of polychromatic plasmonics
and demonstrate the functionalities of a broadband plasmonic lens based on a metal-dielectric-metal (MDM) structure (Fig.~\ref{fig1}). We utilize quadratic modulation of the thickness of the dielectric layer in transverse direction [Fig.~\ref{fig1}(b)] to produce a parabolic optical potential which is practically wavelength independent. We develop analytical descriptions and employ numerical
simulations to show its capability of three-dimensional
subwavelength manipulations, including nanofocusing, self-collimation, and optical pendulum effect. The nanofocusing of
our lens is demonstrated over a bandwidth exceeding an optical
octave ($>500$\,nm) thus allowing for polychromatic plasmon
focusing.

The concept of polychromatic light propagation is well developed for dielectric structures, of which the polychromatic dynamic localization is a noticeable example~\cite{Szameit2009}. In these structures, curved waveguides are employed to produce an effective wavelength invariant optical potential to compensate for the dispersion.

In plasmonics, the engineering of the optical potentials for SPPs has been a key concept for designing plasmonic lenses and other elements~\cite{Yablonovitch2009, Liu2010, Huidobro2010}. Two kinds of structures with transverse~\cite{Verslegers2009, Lee2010} or longitudinally varied effective index~\cite{Stockman2004, Durach2007} are mainly used. In the structures with transverse index modulation the plasmonic potentials are usually heavily wavelength dependent~\cite{Lee2010, Liu2010, Huidobro2010, Verslegers2009}. On the other hand, in longitudinally modulated structures light of different wavelengths can be focused asymptotically~\cite{Durach2007} in the same taper, however light is fully localized there and could not propagate beyond the focus point~\cite{Stockman2004,Durach2007}. Recently a variation of a plasmonic lens was shown to focus broadband light based on diffraction ~\cite{Gao2010}. However, the focusing was obtained for light in free space rather than for propagating SPPs, thus with low focusing resolution. The focusing of polychromatic SPPs still remains an unsolved challenge.

\pict[0.99]{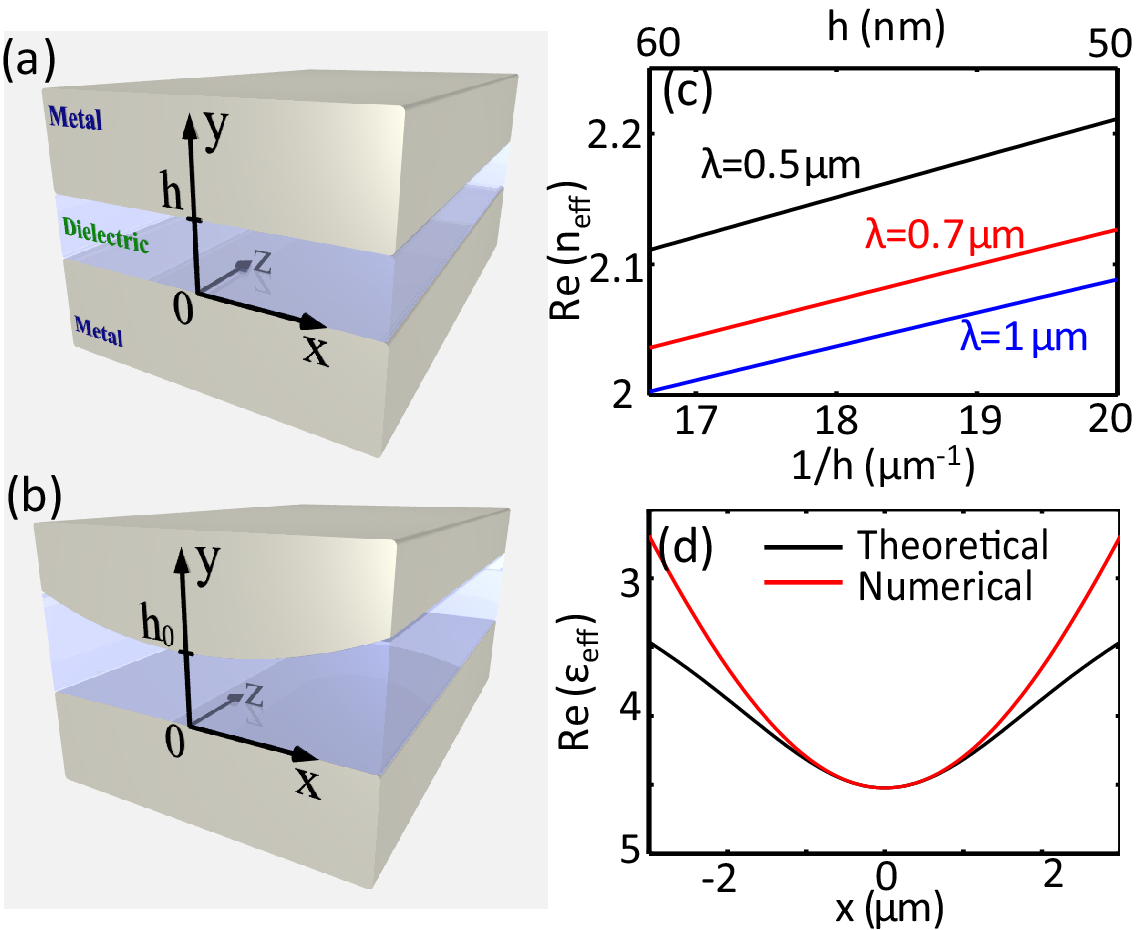}{fig1}{(Color online) (a) Flat and (b) parabolically modulated MDM. (c) Real part of the symmetric mode index of the flat structure vs. $1/h$ for various values of free space wavelengths. 
(d) Real part of the effective permittivity (optical potential) of the structure (b), obtained using Eq.~(\ref{parabolic_pot}) (black) and numerically (red).}

To address this challenge, first we consider a flat MDM waveguide shown in Fig.~\ref{fig1}(a).
We only consider the symmetric modes of the structure (with respect to the magnetic field distribution) because it is primarily excited by simple end-fire coupling, while the antisymmetric mode experiences a cut-off for the parameters of our work~\cite{Zayats2005}.
The dielectric is chosen to be silica glass with permittivity
$\epsilon=2.25$, and the metal is silver for which we use the Drude
model, $\varepsilon_{\rm{m}} = 1 - {\omega_p^2}/(\omega^2 + i\omega
\omega_c)$, where ${\omega}$, ${\omega_p}$, ${\omega_c }$ are the
angular frequency, plasma frequency and collision frequency,
respectively. This model is proven to be a good approximation for
noble metals, including silver in the spectral range above 500\,nm.
For silver we take ${\omega_p}=1.37\times10^{16}$ rad/s and
${\omega_c/\omega_p}=0.002$. Fig.~\ref{fig1}(c) shows the real part
of the mode index, $n_{\rm eff}$, versus the inverse thickness of
the dielectric for three different wavelengths. The dependence is
linear, consistent with the theoretical approximation~\cite{Bozhevolnyi2008,Bozhevolnyi2009},
%
\begin{equation}
\label{n_eff}
n_{\rm eff}  = a/h + b,
\end{equation}
where both $a$ and $b$ are complex parameters and could be extracted
from data fitting. Most importantly, the slope $a$ of these curves is practically {\it wavelength independent} which allows for the design of broadband optical potential and polychromatic plasmon propagation.

Using the effective refractive index in Eq.~(\ref{n_eff}), it is possible to construct a parabolic optical potential~\cite{Arnaud1976} for SPPs. This is achieved by a MDM waveguide with one flat and one parabolically curved surfaces as shown in Fig.~\ref{fig1}(b). The thickness of the dielectric is $h(x) = h_0 + x^2/2R_0$, with $R_0$ as an effective radius ($R_{0}\gg h_{0}, |x|$).
In this waveguide, we obtain a parabolic optical potential under the condition of $x^2 \ll 2h_0 R_0$:
\begin{equation}
\label{parabolic_pot}
\varepsilon_{\rm eff} = n_{\rm eff}^2 (x) = n_0^2 (1 - \Omega^2 x^2),
\end{equation}
where $n_0 = a/h_0 + b$, and $\Omega  = \sqrt{a/(n_0 R_0 h_0^2)}$ is
the focusing strength. Fig.~\ref{fig1}(d) shows the results
calculated by Eq.~(\ref{parabolic_pot}) and numerically for
$h_0=50$\,nm, $\lambda = 0.7\,\mu$m, and $R_0=100\,\mu$m.

Using the effective index method for the structure shown in Fig.~\ref{fig1}(b), we can express the vertical electric field as $E_y (x,y,z) = A(x)B(x,y)\exp(i\beta z)$~\cite{Bozhevolnyi2006,Bozhevolnyi2009}.
When $|x| \ll R_0$, $A(x)$ and $B(x,y)$ could be decoupled~\cite{Bozhevolnyi2009}. The expression
for $B(x,y)$ could be found in Refs.~\cite{Zayats2005,
Bozhevolnyi2009}, while the equation for $A(x)$ is~\cite{Arnaud1976,
Marcuse1972}
\begin{equation}
\label{oscillator}
\frac{d^2 A(x)}{dx^2}+\left[ {\zeta^2 (x) - \beta^2 }\right]A(x) = 0,
\end{equation}
where $\zeta(x)= kn_{\rm eff}(x)$ and $\zeta_0 = kn_0$, $k=2\pi/\lambda$. This is a harmonic oscillator equation with eigenmodes $A_m(x) = {(\sqrt{\pi}{\eta _0}{2^m} m!)^{-\frac{1}{2}}}H_m(x/\eta_0)\exp\left(-x^2/2\eta_0^2\right)$~\cite{Arnaud1976, Marcuse1972}, where $\eta_0 = (\zeta_0 \Omega )^{-\frac{1}{2}}$ is the characteristic width of the plasmonic waveguide, $H_m$ is the Hermite polynomial and the effective indices of different modes, under the paraxial approximation, are $n_{\rm eff}(m) = [\zeta_0-\frac{1}{2}(2m+1)\Omega]/k$.

\pict[0.99]{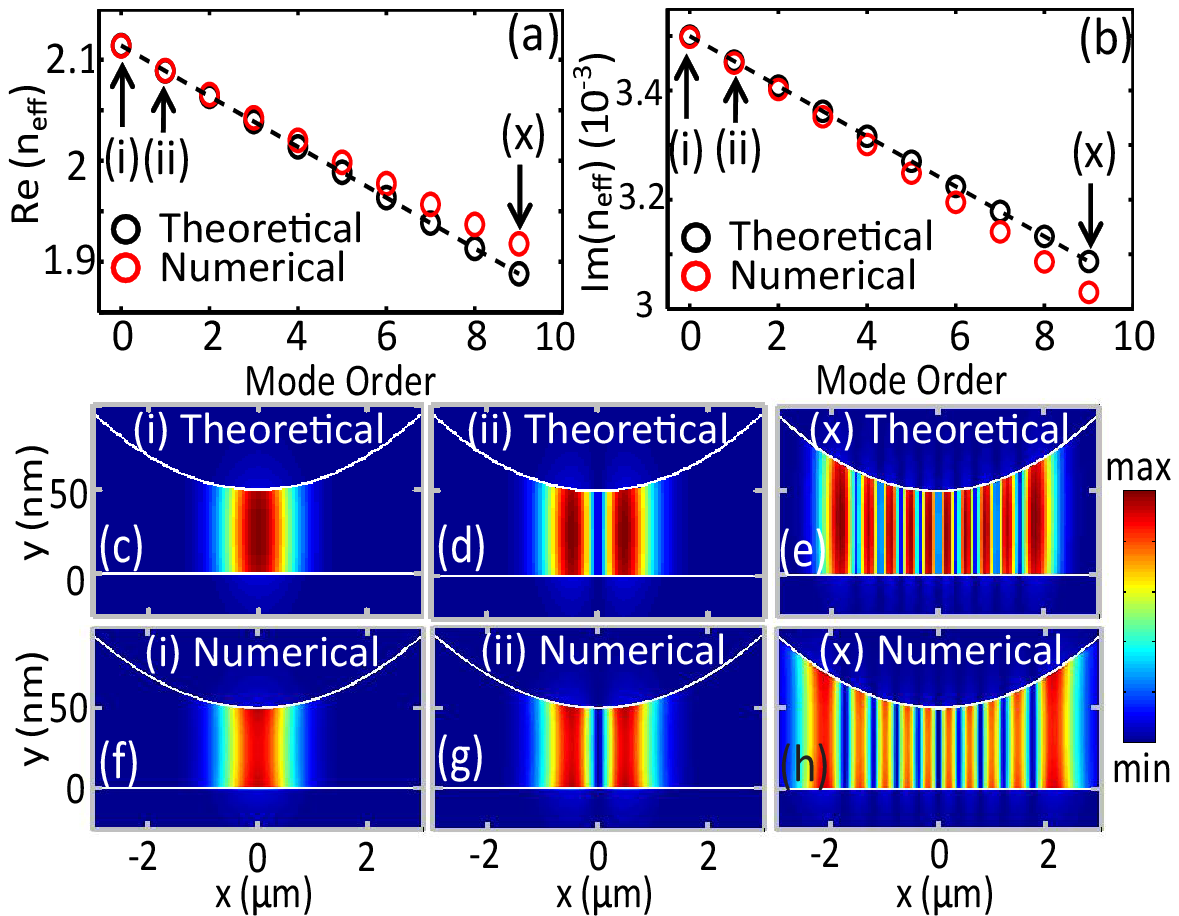}{fig2}{(Color online) (a) Real and (b) imaginary parts of $n_{\rm eff}$
for different order modes. Theoretical (black circles) and numerical (red circles) results are shown.
(c-h) Transverse mode profiles of the fundamental, second and tenth order modes. (c-e) Theoretically and (f-h) numerically calculated mode profiles ($|E_y|$).}

We also calculate the eigenmodes of this structure using commercial
Mode Solutions (MS) software (Lumerical) with the parameters
$x\in[-3, 3]\,\mu$m, $y \in [-1, 1.05]\,\mu$m, $R_0  = 100\,\mu$m, $h_0 = 50$\,nm and $\lambda
= 0.7\,\mu$m. The results of our analytical theory [Eq.~(\ref{oscillator})] and the MS calculations are
summarized in Fig.~\ref{fig2}. Fig.~\ref{fig2}(a,b) shows the real
and imaginary parts of $n_{\rm eff}$ for different modes with the
theoretical field distribution of $|A_m(x)B(x,y)|$
[Fig.~\ref{fig2}(c-e)] and numerically calculated total field $|E_y|$
[Fig.~\ref{fig2}(f-h)], which show a good agreement for low order
modes. For higher order modes, larger discrepancy appears in terms
of both effective index and field distribution. This is consistent with the results in Fig.~\ref{fig1}(d), as the higher order modes spread out to the larger $x$ values, where the assumption $x^2 \ll 2h_0 R_0$ does not strictly hold and the potential is not exactly parabolic.

Next we study the SPPs propagation in the structure. An incident
beam could excite modes of different orders, which will then
interfere with one another, producing different intensity patterns
inside the structure. As a result of this interference, a range of
SPPs beam manipulations is possible, including focusing, self-collimation, and optical pendulum effect~\cite{Arnaud1976, Marcuse1972}.

When $A(x)$ can be decoupled from $B(x,y)$ and $B(x,y)$ changes
slowly with $x$, the input beam expansions along $x$ and $y$ inside the structure are independent. Considering that the beam distribution along $y$ can be fully characterized by $B(x,y)$, we study only the beam dynamics in the $x-z$ plane. The beam could be expanded into a complete set of orthogonal modes~\cite{Marcuse1972} $C(x,z) = \sum\nolimits_{m= 0}^\infty {{a_m}{A_m}(x)\exp (i{\beta_m}z)}$, where $a_m$ is the expansion coefficient of the $m$-order mode. If the initial beam has a Gaussian distribution $C(x,0) = \pi^{-\frac{1}{4}} w_0^{-\frac{1}{2}} \exp[-(x - x_0)^2/2 w_0^2]$, where $w_{0}$ is the beam width, the beam inside the structure, under paraxial approximation, is~\cite{Arnaud1976, Marcuse1972}:
\begin{eqnarray}
\label{eq9}
\begin{array}{l}
 C(x,z) = \pi^{-\frac{1}{4}} q(z)^{-\frac{1}{2}} \exp\left\{-\frac{1}{2\eta^2}\left[ {x-x_0 \cos(\Omega z)} \right]^2
 \right\}\\~~~~~~~~~~~~~\times\exp \left[ i(k_0 z + \frac{1}{2}k_0 \rho^{-1} x^2) \right]P(x_0 ),
 \end{array}
\end{eqnarray}
where $\eta^2 = |q(z)|^2 $,
$\rho^{-1}=\eta^{-1}{d\eta/dz}$, $P(x_0)=\exp \{ i(\eta_0/w_0)^2 [xx_0\sin (\Omega z) - \frac{1}{4}x_0^2 \sin(2\Omega z)]\}$ and $q(z) = w_0 \cos (\Omega_z) + i(\eta _0^2/w_0)\sin (\Omega z)$. In the lossless case with
$\omega_{c}=0$, the beam intensity can be expressed as:
$\left| C(x,z) \right| = \pi^{-\frac{1}{4}} \eta^{-\frac{1}{2}} \exp \left\{ -\left[ x -x_0 \cos(\Omega z) \right]^2 /{2\eta^2} \right\}$ with dynamic beam width varying along propagation:
\begin{eqnarray}
\eta= \sqrt {w_0^2 \cos^2(\Omega z) + \eta _0^4 /w_0^2\sin^2(\Omega z)}\;.
\end{eqnarray}
The plasmon polariton beam is thus oscillating periodically along $z$. It is easy to prove that the maximum and minimum dynamic beam widths are constrained by~\cite{Arnaud1976} $\eta_{\max } \eta_{\min } = \eta _0^2 $, indicating that the beam is trapped in the parabolic optical potential.

\pict[0.99]{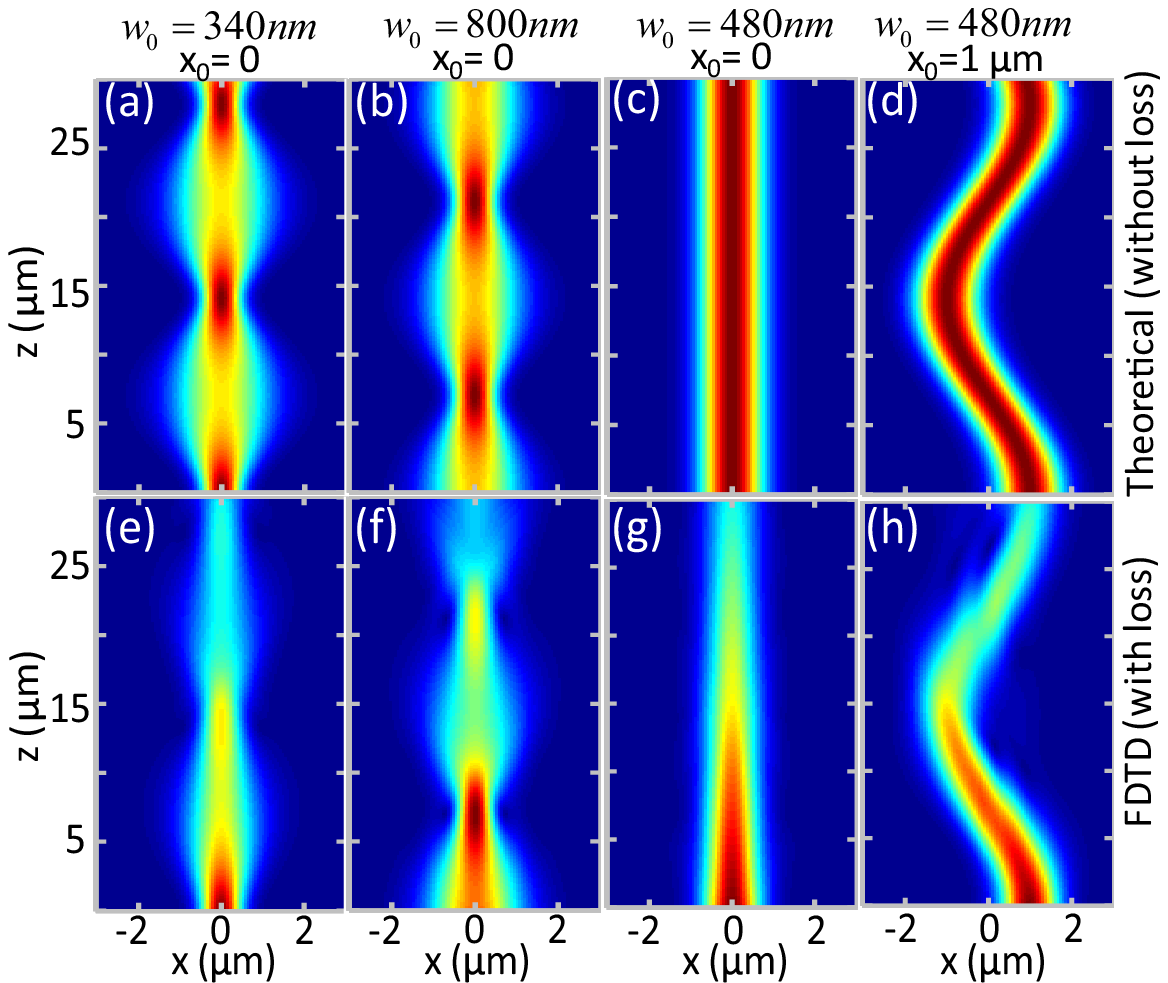}{fig3}{(Color online) (a-d) Theoretical field distribution without loss and (e-h) FDTD simulations with loss (shown $|E|$) in the $x-z$ plane in the middle of the gap ($y=h_0/2=25$\,nm). Plots (a, b, e, f) show typical effects of self-imaging with nanofocusing; (c, g) self-collimation; and (d, h) optical pendulum effect. The corresponding characteristic beam width is $\eta_0\approx480$\,nm.}

To confirm our theoretical analysis, we preform finite-difference-time domain (FDTD) simulations (Lumerical). All initial beams are $y$-polarized with transverse Gaussian distributions to guarantee the excitation of SPPs modes. The structure parameters are the same as in the MS simulation, with $z \in [0,30]\,\mu$m and perfectly matched layers at the boundaries.
The analytical results without loss and the numerical results with loss are presented in Fig.~\ref{fig3}. The data are shown for the
$x-z$ plane in the middle of the gap, $y=h_0/2=25$\,nm.

For an on-axis input beam ($x_{0}=0$) [Fig.~\ref{fig3}(a, b, e, f)], one can observe periodic beam focusing with a period of
\begin{equation}
\label{eq11}
F = \pi/\Omega = \pi\sqrt{R_0h_0(1 +h_0 b/a)}\;.
\end{equation}
The points of maximum intensity correspond to the focusing when all excited modes interfere constructively. There is an initial phase shift depending on the ratio of the incident beam width and the
characteristic width of the plasmonic waveguide $w_0/\eta_0$. When
these two widths match $w_0=\eta_0$, one can observe self-collimation effect - diffractionless propagation of the beam
[see Fig.~\ref{fig3}(c, g)]. For an off-axis input beam ($x_{0}\neq0$)
there are transverse oscillations [see Fig.~\ref{fig3}(d, h)], and
the plasmon polariton beam propagation exhibits an analog of optical
pendulum behavior in a parabolic potential. The comparison with
FDTD results suggests that losses do not affect the beam propagation except for attenuation along $z$ direction.

\pict[0.9]{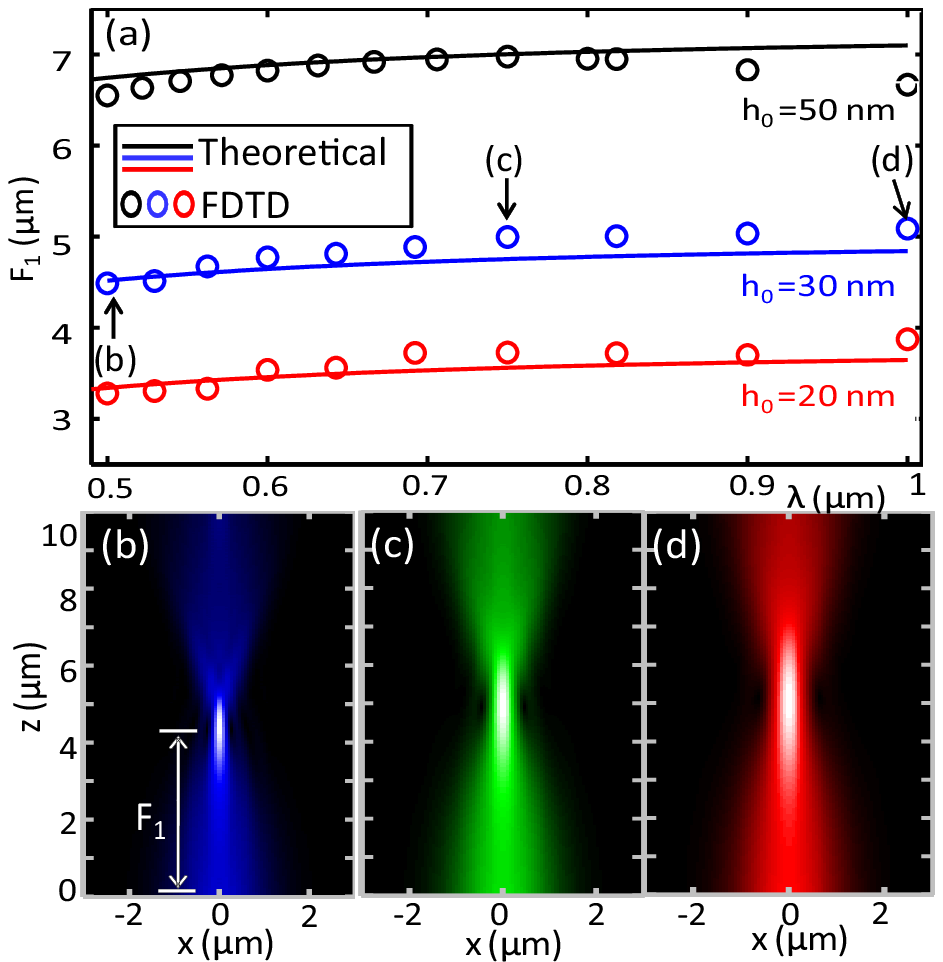}{fig4}{(Color online) (a) Focusing
position $z=F_1$ for wavelengths in the range $0.5-1.0\,\mu$m at
$h_0$ values of 20\,nm, 30\,nm and 50\,nm. Solid lines: theoretical
results; circles: FDTD simulations. (b-d) SPPs propagation for the
three wavelengths marked in (a).}

Eq.~(\ref{eq11}) has some important implications for propagation of
polychromatic SPPs beams. As $a$ is practically wavelength
independent while $b$ varies slowly [Fig.~\rpict{fig1}(c)], the
focusing strengths $\Omega$ and the oscillation period F
are nearly the same for different wavelengths. For example, in
Fig.~\ref{fig4}(a) we show both the theoretical (solid line) and FDTD (circles) results for an input beam with $w_0=800\,{\rm
nm}>\eta_{0}$ for three different thicknesses $h_{0}$ in the wide
wavelength range $0.5-1.0\,\mu$m. The different color beams focus at the planes of $z=F_{1}=F/2$. The dispersionless characteristic of $F$ enables broadband SPPs focusing and distinguishes our plasmonic structure
from other highly wavelength dependent structures~\cite{Lee2010,
Verslegers2009}. In Fig.~\ref{fig4}(b-d) we show the field
distribution $|E|$ on the $x-z$ plane of $y=h_0/2=15$\,nm for the
three points marked in Fig.~\ref{fig4}(a) at $h_0=30$\,nm. False
colors are used to indicate schematically the three different
wavelengths. It is important to note that the focal planes of all
spectral components coincide within 500\,nm (less than one wavelength deviation in the spectral range $0.5-1.0\,\mu$m) and are practically indistinguishable. Note that the focal depth is in the range of $2-3\,\mu$m for the different spectral components. Fig.~\ref{fig4} proves
that this structure acts as a lens for broadband polychromatic light
focusing, a functionality that has never before been discussed in
other plasmonic structures.
From Fig.~\ref{fig4} and Eq.~(\ref{eq11}), it is clear that one can control the position of the focusing points by varying the thicknesses $h_0$ and/or the wire curvature $R_0$, which enables wide reconfigurability of our structure.

\pict[0.99]{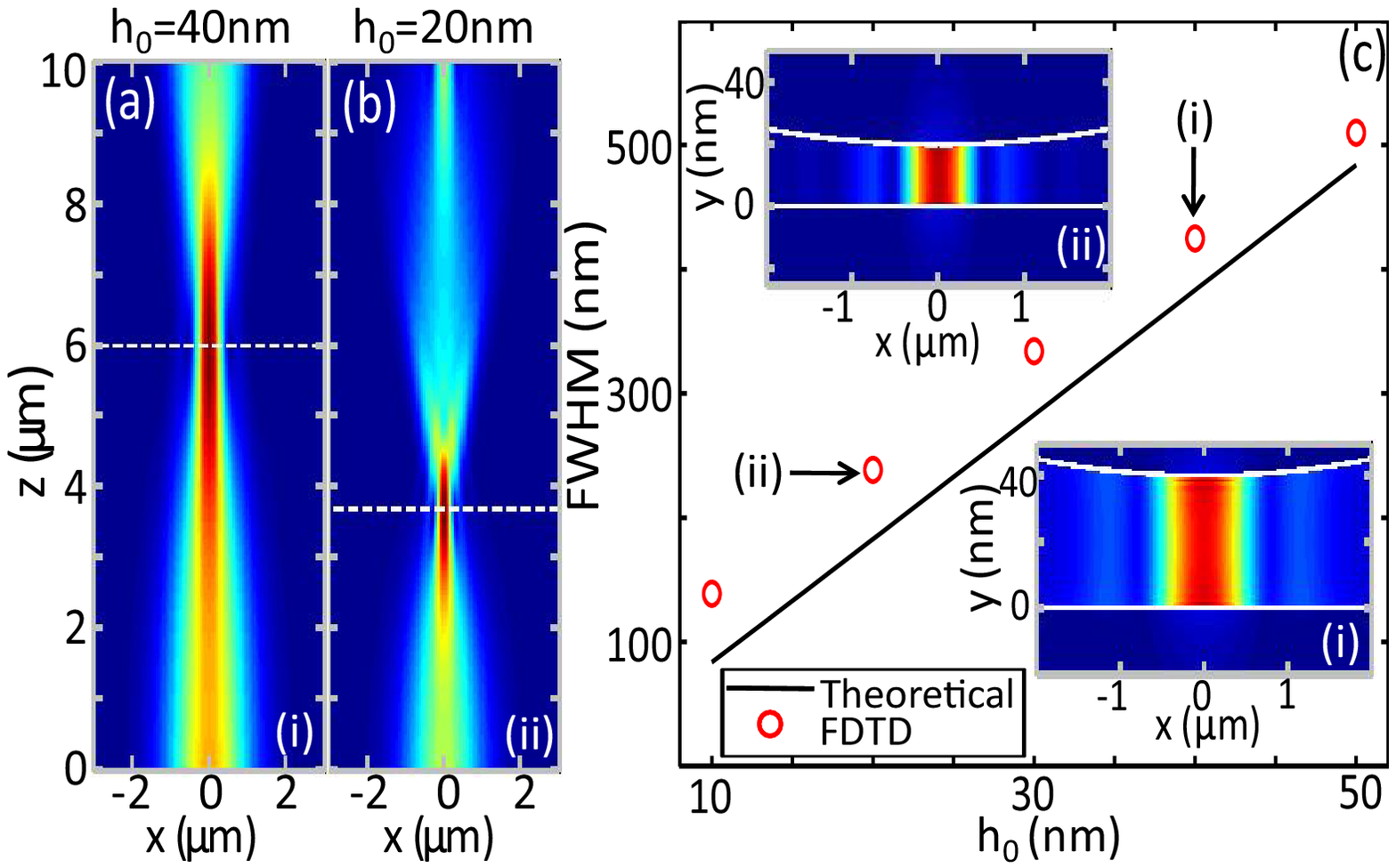}{fig5}{(Color online) (a,b) FDTD results for the field distribution, $|E|$, in the $x-z$ plane $y=h_0/2$ ($h=40$\,nm and 20\,nm) and $\lambda=0.7\,\mu$m. White dashed
lines indicate the focal plane. (c) FWHM in the focal plane
vs. $h_0$. Marked points correspond to the cases (a) and (b),
respectively. Insets: field distributions $|E|$ in the two marked
focal planes.}

Finally, we investigate the focusing resolution of this structure, which is characterized by the full width at half maximum (FWHM) along $x$ direction at the focal plane. (Along the vertical direction the resolution can be approximately characterized by the thickness of the dielectric layer as shown in the insets of Fig.~\ref{fig5}.) We define the wavelength of the SPPs, $\lambda_{\rm spp} = \lambda / n_{\rm eff}$, which is smaller than the light wavelength in the dielectric, $\lambda_p = \lambda/ n_d$. What makes SPPs special is that its wavelength could be made as small as required, even vanishing for localized surface plasmons~\cite{Zayats2005}. In our lens-like structure, if we want to improve the focusing resolution, we could decrease $\lambda_{\rm spp}$. For a fixed vacuum wavelength, the simplest approach to increase $n_{\rm eff}$ is to make the thickness $h_0$ smaller. Fig.~\ref{fig5} shows the dependency of FWHM along $x$ at the focal plane for a fixed wavelength $\lambda=0.7\,\mu$m and $w_0=800$\,nm. Fig.~\ref{fig5}(a, b) shows the field distribution $|E|$ along propagation for the two points marked in Fig.~\ref{fig5}(c) in the $x-z$ plane of $y = h_0/2$ (20\,nm and 10\,nm, respectively). The white dashed lines indicate the focal planes. Insets (i) and (ii) show the transverse field distribution of $|E|$ at the two focusing planes. As shown in Fig.~\ref{fig5}, the FWHM could be as small as 100\,nm. This could be further improved by decreasing $R_{0}$ and/or increasing the focusing strength $\Omega$.
Our results clearly show that the light is fully confined in the dielectric along $y$ and trapped in the optical parabolic potential along $x$.

In conclusion, we have suggested the concept of polychromatic plasmonics and demonstrated a broadband plasmonic lens based on a parabolically modulated MDM structure.
We have shown the plasmonic lens allows for complete three-dimensional subwavelength control of a beam, bringing a potential functionality for manipulation of ultra-short optical signals.
The focusing capability of this plasmonic lens could be further improved by tapering the dielectric layer along the propagation direction~\cite{Bozhevolnyi2010}. We anticipate the proposed structure is a promising candidate for broadband plasmonic
applications including subwavelength white light imaging, polychromatic plasmon
solitons, and ultrashort pulse plasmonic nanolasers. Furthermore, our ideas can be extended beyond the field of plasmonics, to include quantum particles in parabolic potentials, considering the similarity between Shr\"odinger equation and paraxial wave equation.

We thank A.~A. Sukhorukov, Z. Xu, A.~S.~Desyatnikov, C.~G. Poulton, and A.~E. Minovich
for useful discussions, and acknowledge a support from the Australian Research Council
and the NCI
Merit Allocation Scheme.


\end{document}